\documentclass[aps,prd,reprint,groupedaddress,showpacs]{revtex4-1}
\usepackage{amsmath,amssymb,graphicx,caption,subcaption}

\newcommand{\eqn}[1]{\begin{eqnarray} #1 \end{eqnarray}}
\newcommand{\bk}[1]{\langle #1 \rangle}

\newcommand {\ket}[1] {| #1 \rangle}

\begin{document}

\title{Open timelike curves violate Heisenberg's uncertainty principle}

\author{J.L. Pienaar}
 \email{j.pienaar@physics.uq.edu.au}
 \affiliation{
 School of Mathematics and Physics, The University of Queensland, Brisbane 4072 QLD Australia
}
\author{C.R. Myers}
 \email{myers@physics.uq.edu.au}
 \affiliation{Centre for Engineered Quantum Systems, University of Queensland, St. Lucia, QLD 4072, Australia}

\author{T.C. Ralph}
 \email{ralph@physics.uq.edu.au}
\affiliation{
 School of Mathematics and Physics, The University of Queensland, Brisbane 4072 QLD Australia
}

\date{\today}

\begin{abstract}
Toy models for quantum evolution in the presence of closed timelike curves (CTCs) have gained attention in the recent literature due to the strange effects they predict. The circuits that give rise to these effects appear quite abstract and contrived, as they require non-trivial interactions between the future and past which lead to infinitely recursive equations. We consider the special case in which there is no interaction inside the CTC, referred to as an \emph{open} timelike curve (OTC), for which the only local effect is to increase the time elapsed by a clock carried by the system. Remarkably, circuits with access to OTCs are shown to violate Heisenberg's uncertainty principle, allowing perfect state discrimination and perfect cloning of coherent states. The model is extended to wave-packets and smoothly recovers standard quantum mechanics in an appropriate physical limit. The analogy with general relativistic time-dilation suggests that OTCs provide a novel alternative to existing proposals for the behaviour of quantum systems under gravity.

\end{abstract}


\maketitle

\textit{Introduction.} Closed timelike curves (CTCs) are trajectories that form closed loops in time, allowing systems to affect events in their own past. In spite of the paradoxes that this seems to imply, these solutions are consistent with general relativity\cite{GOD49,MOR88}. In contrast, standard relativistic quantum field theory is incompatible with globally non-hyperbolic spacetimes, i.e. those containing CTC's~\cite{BIR82}. Two approaches to bridging this impasse have been discussed in the literature: (i) quantum mechanics modifies general relativity in such a way that CTC's cannot form \cite{HAW92,DES92}; or (ii) nonlinear extensions of quantum mechanics can be found which resolve the paradoxical aspects of CTC's \cite{DEU91,HAR94,POL94,LLO11}. The former program has not yet reached a conclusive result \cite{KIM91,VIS03,EAR09}.

\begin{figure}
        \begin{subfigure}[b]{0.4\textwidth}
                \includegraphics[width=\textwidth]{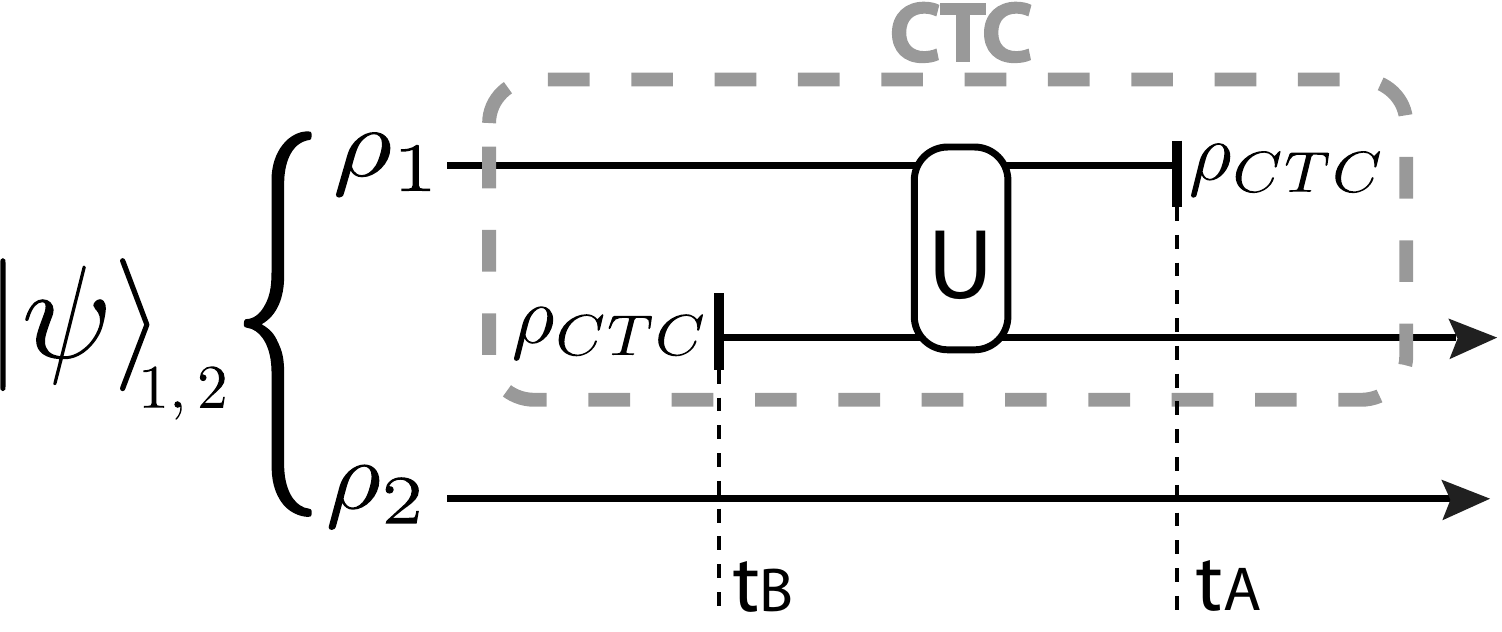}
                \caption{ }
                \label{fig1a}
        \end{subfigure}

        \begin{subfigure}[b]{0.4\textwidth}
                \raggedright
                \includegraphics[width=\textwidth]{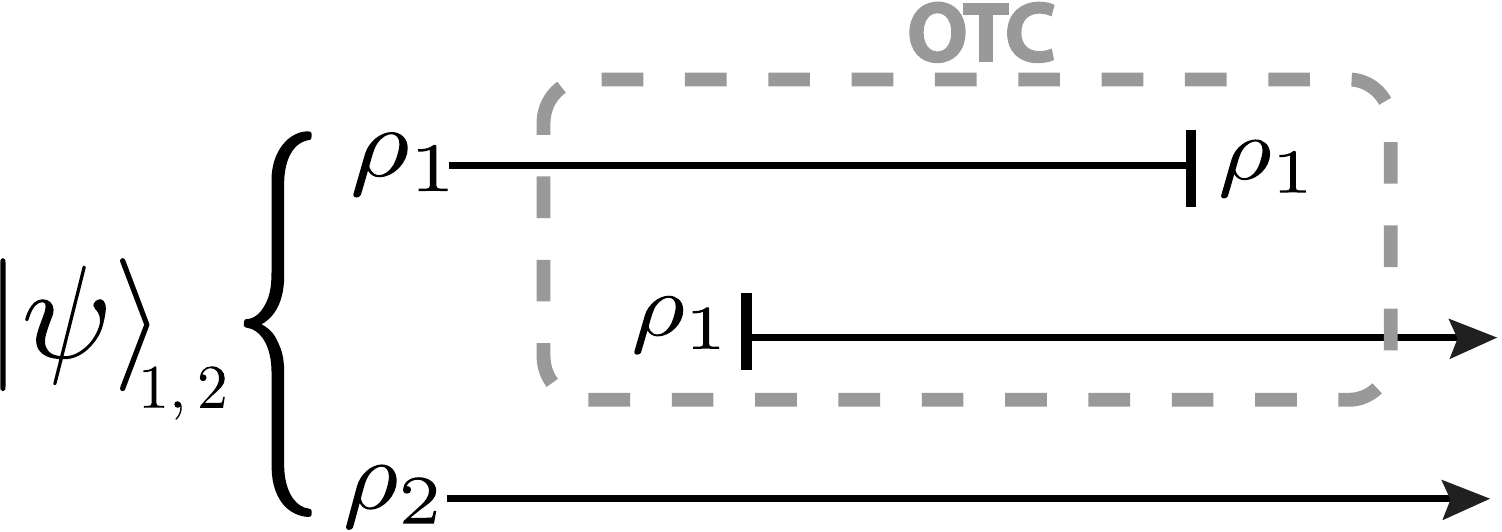}
                \caption{ }
                \label{fig1b}
        \end{subfigure}
        \caption{\raggedright(a) A general closed timelike curve. The time measured by a chronology-respecting system (eg. $\rho_2$) runs horizontally in the figure. A spacetime wormhole \cite{MOR88} is assumed to allow the system in mode 1 (top rail) to jump from point $t_A$ into the past at $t_B$. The possible interaction of the time-travelling system with itself leads to nonlinear effects. (b) The interaction-free case, where $\hat{U}=\hat{I}$. We show that this `open timelike curve' is nevertheless able to violate Heisenberg's uncertainty principle, provided $\rho_1$ is initially entangled with $\rho_2$.\label{fig1} }
\end{figure}

An example of the latter approach is the Deutsch model \cite{DEU91}. In this model, a quantum state is allowed to interact with a version of itself in the past (see Fig.\ref{fig1a}). Such circuits can emulate the famous `grandfather paradox' in which a time-traveller kills their own grandfather, thereby preventing their own existence and leading to a contradiction. For quantum circuits, Deutsch showed that the grandfather paradox is resolved provided the boundary condition between the future and past is enforced by requiring the density operator of the state to match. In particular we require
\begin{equation}\label{SP1}
\rho_{CTC} = Tr_{\ne 1}[U(\rho_{1} \otimes \rho_{CTC})U^{\dagger}]
\end{equation}
where the trace is over all but mode $1$. Given a solution for $\rho_{CTC}$, the output state, $\rho_{out}$, is given by
\begin{equation}\label{SP2}
\rho_{out} = Tr_1[U(\rho_{1} \otimes \rho_{CTC})U^{\dagger}]
\end{equation}
where the trace is now over the Hilbert space of mode 1. Deutsch showed that this procedure leads to solutions for all input states and all unitaries $\hat{U}$ - hence there are no paradoxes. The inherent nonlinearity of Eqs \ref{SP1}, \ref{SP2} leads to unusual properties such as the possibility to build ``super" quantum computers that can solve NP complete problems \cite{BAC04} and the ability to distinguish \cite{BRU09} and clone \cite{AHN10} non-orthogonal qubit states. Some authors have argued that the effects of CTCs are not observable in principle\cite{BEN09}, but this result has been shown not to hold in general\cite{CAV10,CAV12}. In particular, the equivalent circuit interpretation of the Deutsch model avoids this conclusion\cite{RAL10,CAV12}.

At first it may seem that the nonlinearity of the Deutsch model is a direct result of the interaction between the past and present manifestations of the input state. However, in this letter we show that strongly nonlinear effects are still predicted with no interaction in the CTC (i.e. with $\hat{U}=\hat{I}$) provided that entanglement exists between the time-travelling system and an external, chronology respecting system (see Fig.\ref{fig1b}). In particular, we apply the Deutsch model to continuous variable systems and show that the Heisenberg uncertainty principle between canonical variables, such as position and momentum, can be violated in the presence of interaction-free CTCs.

We refer to the interaction-free case as an `open timelike curve' (OTC) because the disconnected paths of the time-traveller's trajectory appear to form an open loop rather than a closed curve. In this case the Deutsch model predicts a time shift on a local clock carried by the system, as well as the decoherence of entanglement. These effects are reminiscent of proposals for nonlinear quantum dynamics in the presence of gravity\cite{RAL09,PEN98}, suggesting a possible connection, which we return to later. We note that in contrast to quantum gravity models which predict an increase in uncertainty due to the discrete nature of spacetime\cite{PIK10}, the model based on OTCs predicts the opposite: namely, the ability to obtain asymptotically \textit{perfect} information about complementary canonical observables. As we will see, this violation of the uncertainty principle then leads to the ability to discriminate and clone unknown states.

\textit{Circuits containing time-travel.} A common feature of all toy models describing quantum evolution in the presence of CTCs is the requirement of nonlinear constraints on quantum mechanics. Early path-integral approaches \cite{HAR94,POL94}, as well as their more recent post-selected equivalents (P-CTCs) \cite{LLO11}, required that entanglement to external systems be preserved. This meant that the consistency conditions had to be applied even to systems in the distant past, leading to the possibility of acausal effects outside the region of the CTCs \cite{HAR94,POL94,RAL12}. While causality violation is expected to occur within the CTC, we might reasonably expect the laws of quantum mechanics to remain causal far away from this region. This is achievable in the Deutsch model, where the constraints apply only to the reduced density matrix of the system at the CTC, thereby breaking the entanglement to external systems. In addition, the model can be interpreted as an equivalent circuit which is free of `information paradoxes'\cite{RAL10}. Given these desirable features, we restrict our attention to the Deutsch model in what follows.

\begin{figure}[!htbp]
 \includegraphics[width=8cm]{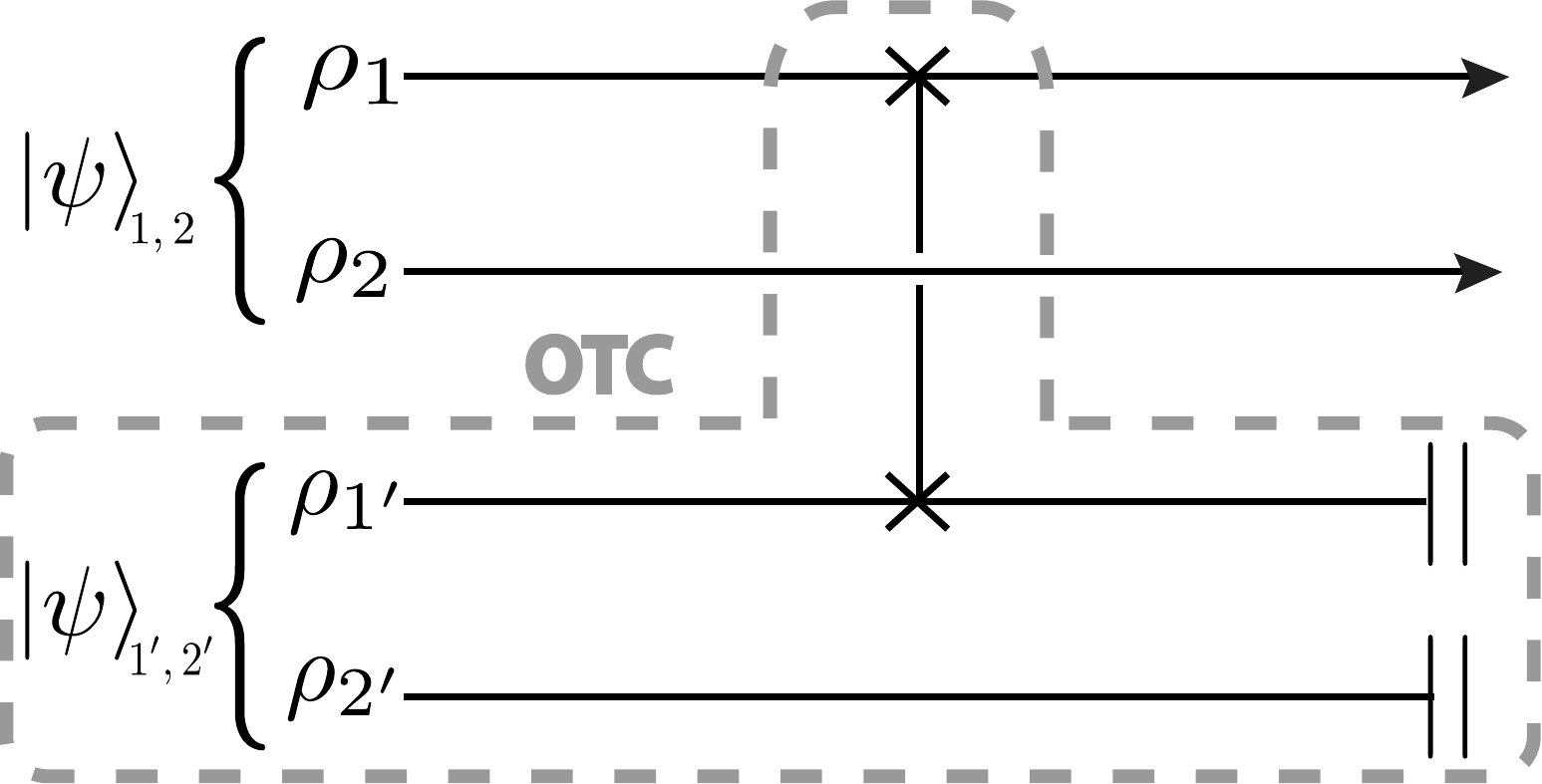}
\caption{\raggedright An equivalent circuit to Fig.\ref{fig1b}. The gate between $\rho_1$ and $\rho_1'$ is a SWAP. This circuit exactly reproduces the input-output relations of an OTC for rails 1 and 2. The lower copy of the circuit enclosed by the dashed line represents degrees of freedom `internal' to the OTC and is therefore inaccessible to outside observers (see \cite{RAL10} for details). \label{fig1c}}
\end{figure} 

The OTC in Fig.\ref{fig1b} can be reproduced by its equivalent circuit, shown in Fig.\ref{fig1c}. This is an ordinary quantum circuit that can be easily constructed in principle, given two copies of the input state. In the case of a real CTC, there is only one available copy of the input state, so the appearance of a second copy represents a violation of the no-cloning theorem\cite{WOO82}, which can be identified as the source of the nonlinearity. In addition, this circuit breaks entanglement to external chronology-respecting systems, which is a necessary feature of the model, as we have discussed. We will soon see how this decoherence effect can be exploited to violate the uncertainty principle. 

\begin{figure}
        \begin{subfigure}[b]{0.4\textwidth}
                \includegraphics[width=\textwidth]{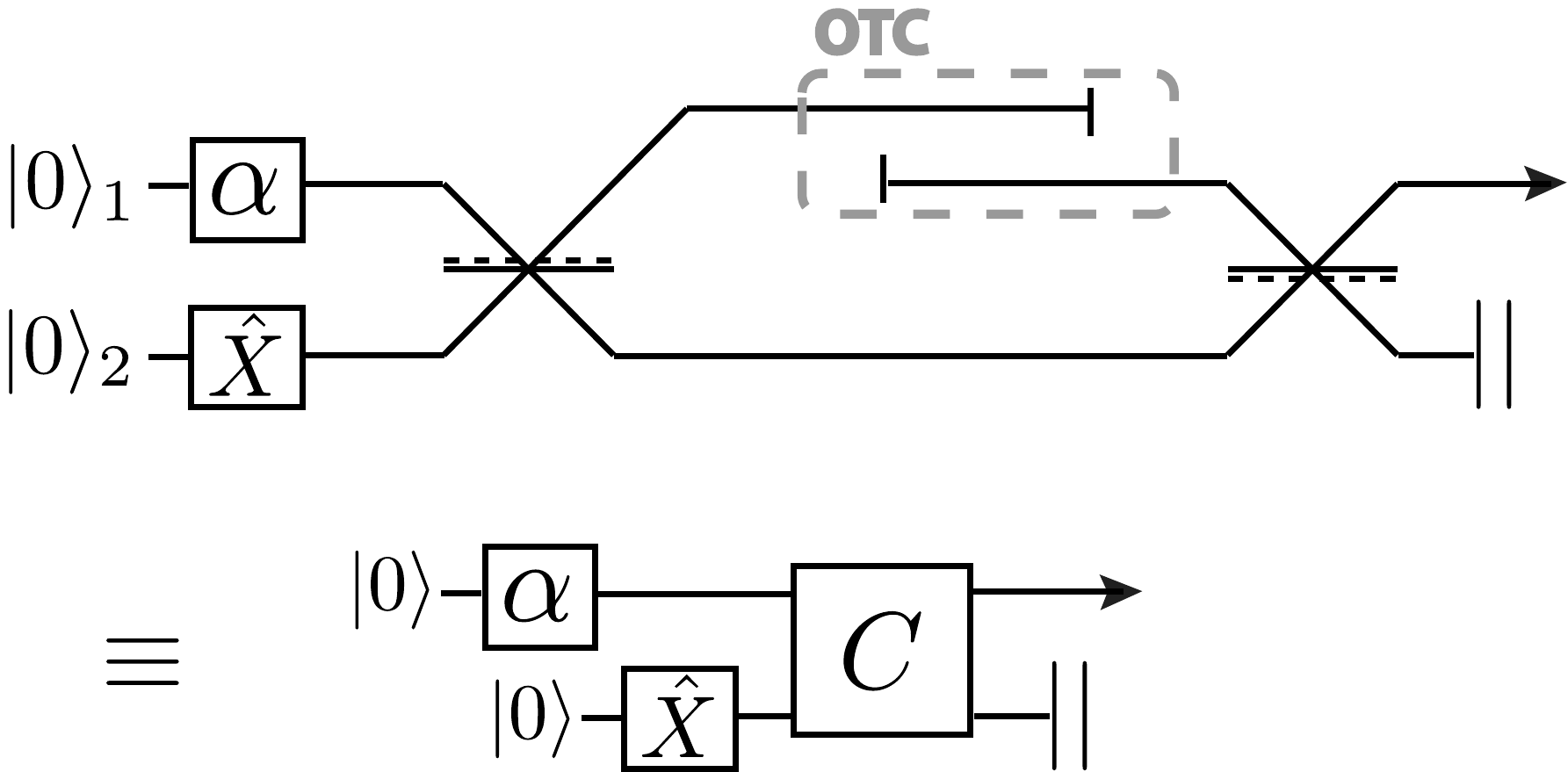}
                \caption{ }
                \label{fig2a}
        \end{subfigure}

        \begin{subfigure}[b]{0.4\textwidth}
                \includegraphics[width=\textwidth]{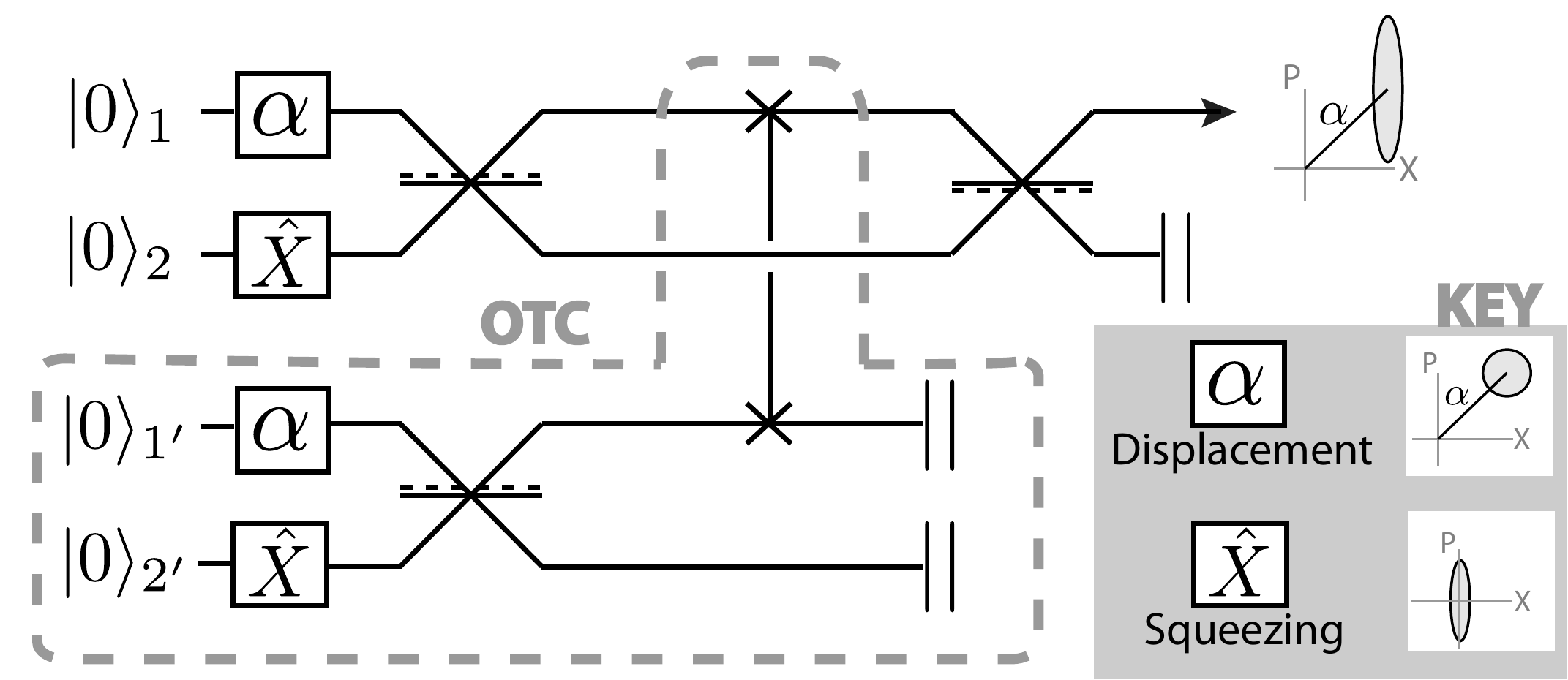}
                \caption{ }
                \label{fig2b}
        \end{subfigure}
        
        \begin{subfigure}[b]{0.4\textwidth}
                \includegraphics[width=\textwidth]{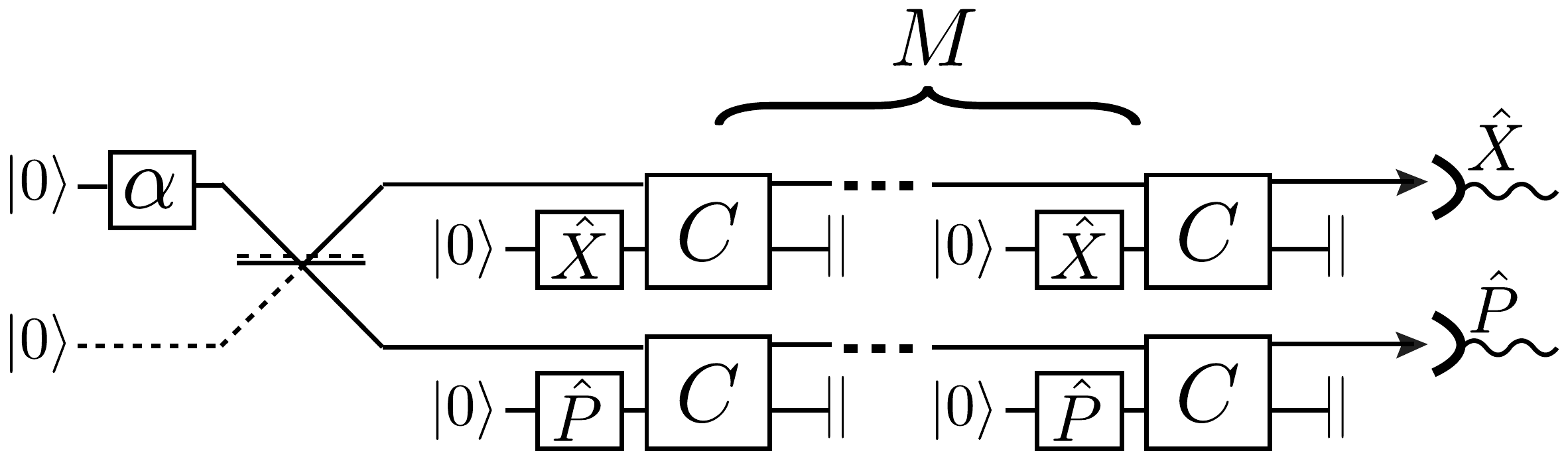}
                \caption{ }
                \label{fig2c}
        \end{subfigure}
        \caption{\raggedright (a) A quantum optics circuit with an OTC on the top rail and (b) its equivalent circuit. Here, the box `$C$' represents coupling to an ancilla that depends upon the input state, which is then traced out. It can therefore be regarded as a nonlinear map on the space of density matrices. (c) A circuit that violates the uncertainty principle. Repeated application of the map $C$ together with appropriately chosen squeezed states leads to arbitrarily precise measurement of $\alpha$ along the orthogonal quadratures $\hat{X},\hat{P}$. \label{fig2}}
\end{figure}

While previous work on CTCs has concentrated almost exclusively on two-dimensional Hilbert spaces, Deutsch's formalism does not restrict the dimensionality of the rails. We are therefore free to interpret each rail as carrying the countably infinite degrees of freedom of a bosonic quantum field.
In the following, for simplicity we will consider a massless scalar field to which we may apply the usual techniques of quantum optics\cite{WAL94}.
Consider the OTC circuit of Fig.\ref{fig2a} and its equivalent circuit shown in Fig.\ref{fig2b}. A coherent state of amplitude $\alpha$ and a squeezed state with squeezing $r$ along the $\hat{X}$ quadrature are prepared on rails 1 and 2 respectively. The rails interact on a 50:50 beamsplitter, rail 1 traverses an OTC, and finally the rails are re-combined on an inverted beamsplitter (note that in the absence of the OTC, the evolution becomes the identity).

Rails 1 and 2 and their copies 1' and 2', hereafter labelled by an abstract index $i \in \{ 1,2,1',2' \}$, are initialised in the field ground state (the vacuum): $\ket{0}_1\ket{0}_2 \ket{0}_{1'} \ket{0}_{2'} \equiv \ket{0}$. The vacuum mode corresponding to rail $i$ is labelled $\hat{a}_i$. The modes $\hat{a}_{1'},\hat{a}_{2'}$ are inaccessible from outside the OTC; hence all physically possible measurements are expectation values on the outputs of $\hat{a}_{1},\hat{a}_{2}$. Let us consider the output $\hat{a}_{1out}$. Using the standard Heisenberg equations of motion for the gates in the circuit, we calculate the output mode in terms of input modes as:
\eqn{\label{aout}  \hat{a}_{1out} &=& \alpha + \frac{1}{2}(\hat{a}_{1} + \hat{a}_{1'})- \frac{i}{2}\, Cosh(r)\, (\hat{a}_{2} - \hat{a}_{2'}) \nonumber \\
&&+ \frac{i}{2}\, Sinh(r)\, (\hat{a}^\dagger_{2} - \hat{a}^\dagger_{2'}) }

We see immediately that $\bk{\hat{a}} = \alpha$, indicating that there is no net displacement in phase space after the evolution. To evaluate the noise properties of the state, consider the quadrature operator $\hat{X}(\theta)=-ie^{i \theta} \hat{a}_{1out}+ie^{-i \theta} \hat{a}^\dagger_{1out}$. The variances along the orthogonal quadratures $\hat{X}=\hat{X}(0)$ and $\hat{P}=\hat{X}(\frac{\pi}{2})$ are exponentially reduced and increased, respectively: $Var\bk{\hat{X}(\frac{\pi}{4}\pm\frac{\pi}{4})} = e^{\pm r} \, Cosh(r)$. For large squeezing, $r\gg1$, the variance of the squeezed quadrature approaches $\frac{1}{2}$. This circuit therefore allows us to deterministically squeeze the quantum noise along the chosen quadrature without changing its displacement, a task that is known to be impossible using standard quantum mechanics. With a straightforward extension of this circuit, we can further violate the uncertainty principle.

\textit{Uncertainty principle violation.} The uncertainty principle states that if $\hat{x},\hat{p}$ are two self-adjoint non-commuting operators representing observables, then the errors in a set of single measurements of these observables must satisfy $\sigma_x \sigma_p \geq \left| \frac{i}{2}[\hat{x},\hat{p}] \right|$, where $\sigma_x, \sigma_p$ are the standard deviations of the measurements (i.e. $\sigma_x = \sqrt{Var\bk{\hat{x}}}=\sqrt{\bk{\hat{x}^2}-\bk{\hat{x}}^2}$) and $[\hat{x},\hat{p}]=\hat{x} \hat{p}-\hat{p} \hat{x}$ is the commutator. The observables $\hat{x},\hat{p}$ are commonly taken to represent position and momentum, whence we obtain $\sigma_x \sigma_p \geq \frac{\hbar}{2}$; however, the uncertainty principle holds for any pair of canonical observables. In quantum optics, for example, the quadrature operators $\hat{X},\hat{P}$ play the analogous role to the position and momentum, satisfying $\sigma_X \sigma_P \geq 1$. We now show that it is possible to arbitrarily violate the uncertainty principle for these observables using the circuit shown in Fig.\ref{fig2c}.

In this new circuit, the input coherent state is first combined with the vacuum on a 50:50 beamsplitter, yielding two copies of the state $\ket{\alpha}\rightarrow \ket{\frac{\alpha}{\sqrt{2}}}_A\ket{\frac{\alpha}{\sqrt{2}}}_B$. We send each of these states through a copy of the original circuit, choosing the squeezing in each case such that $\ket{\frac{\alpha}{\sqrt{2}}}_A$ becomes squeezed in the $\hat{X}$ direction, and $\ket{\frac{\alpha}{\sqrt{2}}}_B$ in the orthogonal $\hat{P}$ direction. Assuming we have repeated access to the OTC, we may cycle each of the outputs back through the circuit for a total of $M$ iterations, using new squeezed states on each run. For simplicity, we assume that these extra squeezed states all have the same squeezing amplitude $r$. At the end of $M$ runs, the displacement on each output will be unchanged, but the variances will be reduced in the chosen quadratures. We then make measurements of $\hat{X}_A$ and $\hat{P}_B$ on the respective outputs, and thereby determine the components of $\frac{\alpha}{\sqrt{2}}$ along orthogonal quadrature directions.
The errors for these measurements are equal to:
\eqn{Var\bk{\hat{X}_A}=Var \bk{\hat{P}_B} = \frac{1+2^{M-R}-2^{-R} }{2^{M}} }
where $R \equiv \frac{2}{log(2)} r$ is introduced for clarity. According to Heisenberg's uncertainty principle, the errors in these measurements should satisfy $Var\bk{\hat{X}_A} Var \bk{\hat{P}_B} \geq 1$. However, we see that for $r \gg M$, their product scales as $\frac{1}{2^{M}}$; thus we have measured the orthogonal components of $\alpha$ to an arbitrary degree of accuracy forbidden by the uncertainty principle. This information can then be used to identify and distinguish unknown coherent states and make clones of them as desired. The circuit functions effectively as a nonlinear `state read-out box' for coherent states, since it produces a read-out of the components of $\alpha$ to arbitrary accuracy in a single shot, requiring no prior knowledge of the state (for a general definition of such devices, see\cite{KEN05}). We note that even for a single use of the OTC (i.e. $M=1$) the Heisenberg principle is violated at sufficiently large squeezing.

It has recently been claimed that any theory which violates the Heisenberg uncertainty principle must also violate the second law of thermodynamics\cite{HAN12}. This claim seems to be at odds with Deutsch's original proof that the input-output relations of CTCs are consistent with the second law\cite{DEU91}. The issue is resolved by noting that the authors of \cite{HAN12} assume that measurement effects are \textit{linear} functionals on the state space. Since the input-output maps of the circuit Fig.\ref{fig2c} (and of Deutsch CTCs in general) can be regarded as nonlinear functionals of the input state, this assumption is not satisfied by these models, and therefore can still be thermodynamically consistent.

Finally, we note that this circuit is not sufficient to solve NP-complete problems as has been demonstrated for more general CTCs\cite{BAC04}. The reason is that as we obtain greater accuracy by increasing $M$, we also require an increase in the minimum squeezing $r$ on each squeezed state. This translates into requiring a number of photons that scales exponentially with $M$. While it may not be possible to obtain a computational advantage by considering Gaussian states and operations alone, we conjecture that OTCs will lead to an increase in computational power for non-Gaussian states.

\begin{figure}[!htbp]
 \includegraphics[width=6cm]{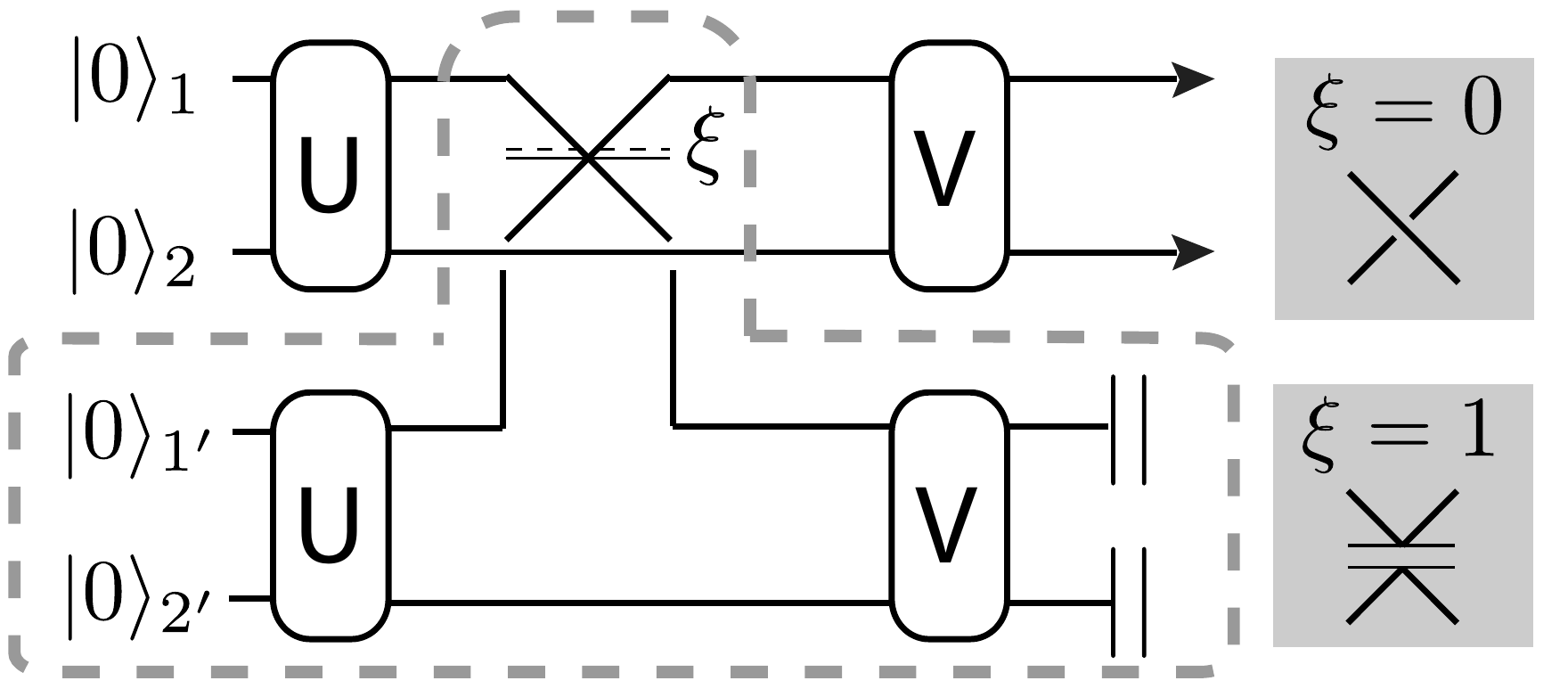}
\caption{\raggedright A generalized equivalent circuit. In the limit $\xi=0$, the beamsplitter functions as a swap gate, leading to the equivalent circuit for an OTC (compare to Fig.\ref{fig1c}). In the opposite limit $\xi=1$, the circuits decouple into two copies of the same circuit, the lower copy being unobservable and the upper copy recovering the standard theory. \label{fig3}}
\end{figure}

\textit{Curvature as an OTC effect.} While the Deutsch model may apply to causality-violating spacetimes, it is generally accepted that standard quantum field theory applies in more well-behaved (globally hyperbolic) curved metrics. However, the lack of experimental data in this regime leaves open the possibility for alternative models, whose predictions may diverge from ordinary quantum mechanics in the presence of gravitational curvature. A field theory that is consistent with the Deutsch model might provide one such candidate. We note that the example of an OTC bears some resemblance to general relativistic time dilation, since it introduces a time difference between two initially synchronised clocks. In particular, an experimenter presented with two such systems could not tell whether the time difference was due to gravitational curvature or to the traversal of an OTC. If we model spacetime curvature as an OTC, this would imply additional decoherence and nonlinear effects not predicted by standard quantum field theory. Nevertheless, we would expect both theories to agree in certain limits such as flat spacetime. To make this idea more concrete, we generalise the equivalent circuit to recover linear quantum mechanics. This is achieved by replacing the swap gate with a beamsplitter with reflectivity $\xi$ (see Fig.\ref{fig3}).

We see that for $\xi=0$ we obtain the circuit of an OTC, but when $\xi=1$ we recover standard quantum theory. By varying the parameter $\xi$ we can smoothly transition between these two limiting cases. All that remains is to define $\xi$ in terms of physical parameters. 

To this end, consider the detection of two entangled wave-packets propagating through a spacetime containing an OTC, such that one trajectory passes through the OTC while the other does not\footnote{One example of an appropriate metric can be found in \cite{POL94}.}. The field annihilation operators for the two wave-packets at the detectors are denoted $\hat{a}_i \rightarrow \int \mathrm{d}\textbf{k} \; G_i(k,x) \hat{a}_{i,\textbf{k}} \equiv \hat{A}_{G,i}(x)$, where $i=1,2$. The function $G_i(k,x)$ is a normalised solution of the Klein-Gordon equation with $k,x$ being the momentum and position four-vectors satisfying $kx = g_{\mu \nu}k^{\mu}x^{\nu}$ and $g_{\mu \nu}$ provides the local metric along the trajectories. The temporal resolution of the detectors is given by the functions $G_i(\tau)$\footnote{More precisely, we consider the time co-ordinate $\tau$ in the reference frame of the detectors, which are assumed to be at rest relative to each other. The spatial wave-packet on a given time-slice $\tau=t$ is denoted $G_i(\textbf{x},t)$. If $(\textbf{x}_{i},\tau')$ is the detection event for the $i_{th}$ trajectory, then the temporal detector resolution is $G_i(\textbf{x}_i,\tau-\tau')\equiv G_i(\tau)$. See \cite{RAL09,PIE11} for details.}. We postulate the parameter $\xi$ measures how well, in principle, the detectors could resolve the time difference $\Delta T$, via the normalised overlap:
\eqn{\label{overlap} \xi \equiv \frac{\int \mathrm{d} \tau G_1^{*}(\tau+\Delta T) G_2(\tau) }{\int \mathrm{d} \tau G_1^{*}(\tau) G_2(\tau) } .}
Assuming the wave-packets to be approximately Gaussian in shape, we see that when $\Delta T$ is large compared to the coherence time of the wave-packets (the width of $G_i(\tau)$), then $\xi \rightarrow 0$ and we obtain the equivalent circuit for an OTC. Conversely, when $\Delta T$ is much smaller than the coherence time, $\xi \rightarrow 1$ and we recover the predictions of the standard theory. 

We have assumed that the time difference between the two trajectories was caused by the presence of an OTC in the metric. However, as argued above, we could equally well apply this model to any instance of gravitational time-dilation, as an alternative to the standard paradigm of quantum field theory. In fact, this interpretation reproduces the nonlinear `event operator' formalism for optical fields in curved spacetime, proposed in earlier work \cite{RAL09}, at the same time making the formalism more transparent by casting it in equivalent circuit form. If we adopt this interpretation then the model predicts that nonlinear effects could be observable from experiments carried out in the earth's gravitational field\cite{RAL09}. As we have seen, the nonlinear effects may accumulate over multiple traversals of the OTC, which opens up the possibility of more easily attainable experiments; we leave these considerations to future work.

\textit{Conclusions.} We have shown that OTCs, despite being the simplest type of CTC interaction, nevertheless are able to violate the Heisenberg uncertainty principle for continuous variable systems. Specifically, we showed that a simple circuit involving linear optics and repeated access to an OTC is sufficient to arbitrarily well distinguish non-orthogonal coherent states and to clone them.

Since CTCs are predicted by general relativity, it is reasonable to ask whether the laws of physics can accommodate these solutions without leading to paradoxes. This can be achieved by making quantum mechanics nonlinear, as in the Deutsch model. The result of this paper indicates that the simple requirement that quantum theory be consistent with time travel leads to highly nonlinear effects, regardless of any interaction on the CTC. We have speculated on the extension of this idea to curved spacetimes in general by conjecturing that gravitational time dilation can be modelled as an OTC effect. This leads to a modified theory of quantum optics that becomes nonlinear in the presence of curvature. This speculation could be testable with current technology via experiments on entangled systems in earth's gravitational field.

\begin{acknowledgements}
The authors thank Joshua Combes for interesting discussions.
\end{acknowledgements}

\bibliographystyle{apsrev4-1}

\begin{thebibliography}{30}%
\makeatletter
\providecommand \@ifxundefined [1]{%
 \@ifx{#1\undefined}
}%
\providecommand \@ifnum [1]{%
 \ifnum #1\expandafter \@firstoftwo
 \else \expandafter \@secondoftwo
 \fi
}%
\providecommand \@ifx [1]{%
 \ifx #1\expandafter \@firstoftwo
 \else \expandafter \@secondoftwo
 \fi
}%
\providecommand \natexlab [1]{#1}%
\providecommand \enquote  [1]{``#1''}%
\providecommand \bibnamefont  [1]{#1}%
\providecommand \bibfnamefont [1]{#1}%
\providecommand \citenamefont [1]{#1}%
\providecommand \href@noop [0]{\@secondoftwo}%
\providecommand \href [0]{\begingroup \@sanitize@url \@href}%
\providecommand \@href[1]{\@@startlink{#1}\@@href}%
\providecommand \@@href[1]{\endgroup#1\@@endlink}%
\providecommand \@sanitize@url [0]{\catcode `\\12\catcode `\$12\catcode
  `\&12\catcode `\#12\catcode `\^12\catcode `\_12\catcode `\%12\relax}%
\providecommand \@@startlink[1]{}%
\providecommand \@@endlink[0]{}%
\providecommand \url  [0]{\begingroup\@sanitize@url \@url }%
\providecommand \@url [1]{\endgroup\@href {#1}{\urlprefix }}%
\providecommand \urlprefix  [0]{URL }%
\providecommand \Eprint [0]{\href }%
\@ifxundefined \urlstyle {%
  \providecommand \doi  [0]{\begingroup \@sanitize@url \@doi}%
  \providecommand \@doi [1]{\endgroup \@@startlink {\doibase
  #1}doi:\discretionary {}{}{}#1\@@endlink }%
}{%
  \providecommand \doi  [0]{doi:\discretionary{}{}{}\begingroup
  \urlstyle{rm}\Url }%
}%
\providecommand \doibase [0]{http://dx.doi.org/}%
\providecommand \Doi [0]{\begingroup \@sanitize@url \@Doi }%
\providecommand \@Doi  [1]{\endgroup\@@startlink{\doibase#1}\@@Doi}%
\providecommand \@@Doi [1]{#1\@@endlink}%
\providecommand \selectlanguage [0]{\@gobble}%
\providecommand \bibinfo  [0]{\@secondoftwo}%
\providecommand \bibfield  [0]{\@secondoftwo}%
\providecommand \translation [1]{[#1]}%
\providecommand \BibitemOpen [0]{}%
\providecommand \bibitemStop [0]{}%
\providecommand \bibitemNoStop [0]{.\EOS\space}%
\providecommand \EOS [0]{\spacefactor3000\relax}%
\providecommand \BibitemShut  [1]{\csname bibitem#1\endcsname}%
\bibitem [{\citenamefont {G\"odel}(1949)}]{GOD49}%
  \BibitemOpen
  \bibfield  {author} {\bibinfo {author} {\bibfnamefont {K.}~\bibnamefont
  {G\"odel}},\ }\Doi {10.1103/RevModPhys.21.447} {\bibfield  {journal}
  {\bibinfo  {journal} {Rev. Mod. Phys.},\ }\textbf {\bibinfo {volume} {21}},\
  \bibinfo {pages} {447} (\bibinfo {year} {1949})}\BibitemShut {NoStop}%
\bibitem [{\citenamefont {Morris}\ \emph {et~al.}(1988)\citenamefont {Morris},
  \citenamefont {Thorne},\ and\ \citenamefont {Yurtsever}}]{MOR88}%
  \BibitemOpen
  \bibfield  {author} {\bibinfo {author} {\bibfnamefont {M.~S.}\ \bibnamefont
  {Morris}}, \bibinfo {author} {\bibfnamefont {K.~S.}\ \bibnamefont {Thorne}},
  \ and\ \bibinfo {author} {\bibfnamefont {U.}~\bibnamefont {Yurtsever}},\
  }\Doi {10.1103/PhysRevLett.61.1446} {\bibfield  {journal} {\bibinfo
  {journal} {Phys. Rev. Lett.},\ }\textbf {\bibinfo {volume} {61}},\ \bibinfo
  {pages} {1446} (\bibinfo {year} {1988})}\BibitemShut {NoStop}%
\bibitem [{\citenamefont {Birrell}\ and\ \citenamefont {Davies}(1982)}]{BIR82}%
  \BibitemOpen
  \bibfield  {author} {\bibinfo {author} {\bibfnamefont {N.~D.}\ \bibnamefont
  {Birrell}}\ and\ \bibinfo {author} {\bibfnamefont {P.~C.~W.}\ \bibnamefont
  {Davies}},\ }\href@noop {} {\emph {\bibinfo {title} {Quantum Fields in Curved
  Space}}}\ (\bibinfo  {publisher} {Cambridge University Press, Cambridge,
  England},\ \bibinfo {year} {1982})\BibitemShut {NoStop}%
\bibitem [{\citenamefont {Hawking}(1992)}]{HAW92}%
  \BibitemOpen
  \bibfield  {author} {\bibinfo {author} {\bibfnamefont {S.~W.}\ \bibnamefont
  {Hawking}},\ }\Doi {10.1103/PhysRevD.46.603} {\bibfield  {journal} {\bibinfo
  {journal} {Phys. Rev. D},\ }\textbf {\bibinfo {volume} {46}},\ \bibinfo
  {pages} {603} (\bibinfo {year} {1992})}\BibitemShut {NoStop}%
\bibitem [{\citenamefont {Deser}\ \emph {et~al.}(1992)\citenamefont {Deser},
  \citenamefont {Jackiw},\ and\ \citenamefont {'t~Hooft}}]{DES92}%
  \BibitemOpen
  \bibfield  {author} {\bibinfo {author} {\bibfnamefont {S.}~\bibnamefont
  {Deser}}, \bibinfo {author} {\bibfnamefont {R.}~\bibnamefont {Jackiw}}, \
  and\ \bibinfo {author} {\bibfnamefont {G.}~\bibnamefont {'t~Hooft}},\ }\Doi
  {10.1103/PhysRevLett.68.267} {\bibfield  {journal} {\bibinfo  {journal}
  {Phys. Rev. Lett.},\ }\textbf {\bibinfo {volume} {68}},\ \bibinfo {pages}
  {267} (\bibinfo {year} {1992})}\BibitemShut {NoStop}%
\bibitem [{\citenamefont {Deutsch}(1991)}]{DEU91}%
  \BibitemOpen
  \bibfield  {author} {\bibinfo {author} {\bibfnamefont {D.}~\bibnamefont
  {Deutsch}},\ }\Doi {10.1103/PhysRevD.44.3197} {\bibfield  {journal} {\bibinfo
   {journal} {Phys. Rev. D},\ }\textbf {\bibinfo {volume} {44}},\ \bibinfo
  {pages} {3197} (\bibinfo {year} {1991})}\BibitemShut {NoStop}%
\bibitem [{\citenamefont {Hartle}(1994)}]{HAR94}%
  \BibitemOpen
  \bibfield  {author} {\bibinfo {author} {\bibfnamefont {J.~B.}\ \bibnamefont
  {Hartle}},\ }\Doi {10.1103/PhysRevD.49.6543} {\bibfield  {journal} {\bibinfo
  {journal} {Phys. Rev. D},\ }\textbf {\bibinfo {volume} {49}},\ \bibinfo
  {pages} {6543} (\bibinfo {year} {1994})}\BibitemShut {NoStop}%
\bibitem [{\citenamefont {Politzer}(1994)}]{POL94}%
  \BibitemOpen
  \bibfield  {author} {\bibinfo {author} {\bibfnamefont {H.~D.}\ \bibnamefont
  {Politzer}},\ }\Doi {10.1103/PhysRevD.49.3981} {\bibfield  {journal}
  {\bibinfo  {journal} {Phys. Rev. D},\ }\textbf {\bibinfo {volume} {49}},\
  \bibinfo {pages} {3981} (\bibinfo {year} {1994})}\BibitemShut {NoStop}%
\bibitem [{\citenamefont {Lloyd}\ \emph {et~al.}(2011)\citenamefont {Lloyd},
  \citenamefont {Maccone}, \citenamefont {Garcia-Patron}, \citenamefont
  {Giovannetti}, \citenamefont {Shikano}, \citenamefont {Pirandola},
  \citenamefont {Rozema}, \citenamefont {Darabi}, \citenamefont {Soudagar},
  \citenamefont {Shalm},\ and\ \citenamefont {Steinberg}}]{LLO11}%
  \BibitemOpen
  \bibfield  {author} {\bibinfo {author} {\bibfnamefont {S.}~\bibnamefont
  {Lloyd}}, \bibinfo {author} {\bibfnamefont {L.}~\bibnamefont {Maccone}},
  \bibinfo {author} {\bibfnamefont {R.}~\bibnamefont {Garcia-Patron}}, \bibinfo
  {author} {\bibfnamefont {V.}~\bibnamefont {Giovannetti}}, \bibinfo {author}
  {\bibfnamefont {Y.}~\bibnamefont {Shikano}}, \bibinfo {author} {\bibfnamefont
  {S.}~\bibnamefont {Pirandola}}, \bibinfo {author} {\bibfnamefont {L.~A.}\
  \bibnamefont {Rozema}}, \bibinfo {author} {\bibfnamefont {A.}~\bibnamefont
  {Darabi}}, \bibinfo {author} {\bibfnamefont {Y.}~\bibnamefont {Soudagar}},
  \bibinfo {author} {\bibfnamefont {L.~K.}\ \bibnamefont {Shalm}}, \ and\
  \bibinfo {author} {\bibfnamefont {A.~M.}\ \bibnamefont {Steinberg}},\ }\Doi
  {10.1103/PhysRevLett.106.040403} {\bibfield  {journal} {\bibinfo  {journal}
  {Phys. Rev. Lett.},\ }\textbf {\bibinfo {volume} {106}},\ \bibinfo {pages}
  {040403} (\bibinfo {year} {2011})}\BibitemShut {NoStop}%
\bibitem [{\citenamefont {Kim}\ and\ \citenamefont {Thorne}(1991)}]{KIM91}%
  \BibitemOpen
  \bibfield  {author} {\bibinfo {author} {\bibfnamefont {S.-W.}\ \bibnamefont
  {Kim}}\ and\ \bibinfo {author} {\bibfnamefont {K.~S.}\ \bibnamefont
  {Thorne}},\ }\Doi {10.1103/PhysRevD.43.3929} {\bibfield  {journal} {\bibinfo
  {journal} {Phys. Rev. D},\ }\textbf {\bibinfo {volume} {43}},\ \bibinfo
  {pages} {3929} (\bibinfo {year} {1991})}\BibitemShut {NoStop}%
\bibitem [{\citenamefont {Visser}\ \emph {et~al.}(2003)\citenamefont {Visser},
  \citenamefont {Kar},\ and\ \citenamefont {Dadhich}}]{VIS03}%
  \BibitemOpen
  \bibfield  {author} {\bibinfo {author} {\bibfnamefont {M.}~\bibnamefont
  {Visser}}, \bibinfo {author} {\bibfnamefont {S.}~\bibnamefont {Kar}}, \ and\
  \bibinfo {author} {\bibfnamefont {N.}~\bibnamefont {Dadhich}},\ }\Doi
  {10.1103/PhysRevLett.90.201102} {\bibfield  {journal} {\bibinfo  {journal}
  {Phys. Rev. Lett.},\ }\textbf {\bibinfo {volume} {90}},\ \bibinfo {pages}
  {201102} (\bibinfo {year} {2003})}\BibitemShut {NoStop}%
\bibitem [{\citenamefont {Earman}\ \emph {et~al.}(2009)\citenamefont {Earman},
  \citenamefont {Smeenk},\ and\ \citenamefont {Wüthrich}}]{EAR09}%
  \BibitemOpen
  \bibfield  {author} {\bibinfo {author} {\bibfnamefont {J.}~\bibnamefont
  {Earman}}, \bibinfo {author} {\bibfnamefont {C.}~\bibnamefont {Smeenk}}, \
  and\ \bibinfo {author} {\bibfnamefont {C.}~\bibnamefont {Wüthrich}},\ }\href
  {http://dx.doi.org/10.1007/s11229-008-9338-2} {\bibfield  {journal} {\bibinfo
   {journal} {Synthese},\ }\textbf {\bibinfo {volume} {169}},\ \bibinfo {pages}
  {91} (\bibinfo {year} {2009})},\ ISSN \bibinfo {issn} {0039-7857},\ \bibinfo
  {note} {10.1007/s11229-008-9338-2}\BibitemShut {NoStop}%
\bibitem [{\citenamefont {Bacon}(2004)}]{BAC04}%
  \BibitemOpen
  \bibfield  {author} {\bibinfo {author} {\bibfnamefont {D.}~\bibnamefont
  {Bacon}},\ }\Doi {10.1103/PhysRevA.70.032309} {\bibfield  {journal} {\bibinfo
   {journal} {Phys. Rev. A},\ }\textbf {\bibinfo {volume} {70}},\ \bibinfo
  {pages} {032309} (\bibinfo {year} {2004})}\BibitemShut {NoStop}%
\bibitem [{\citenamefont {Brun}\ \emph {et~al.}(2009)\citenamefont {Brun},
  \citenamefont {Harrington},\ and\ \citenamefont {Wilde}}]{BRU09}%
  \BibitemOpen
  \bibfield  {author} {\bibinfo {author} {\bibfnamefont {T.~A.}\ \bibnamefont
  {Brun}}, \bibinfo {author} {\bibfnamefont {J.}~\bibnamefont {Harrington}}, \
  and\ \bibinfo {author} {\bibfnamefont {M.~M.}\ \bibnamefont {Wilde}},\ }\Doi
  {10.1103/PhysRevLett.102.210402} {\bibfield  {journal} {\bibinfo  {journal}
  {Phys. Rev. Lett.},\ }\textbf {\bibinfo {volume} {102}},\ \bibinfo {pages}
  {210402} (\bibinfo {year} {2009})}\BibitemShut {NoStop}%
\bibitem [{\citenamefont {Ahn}\ \emph {et~al.}(2010)\citenamefont {Ahn},
  \citenamefont {Ralph},\ and\ \citenamefont {Mann}}]{AHN10}%
  \BibitemOpen
  \bibfield  {author} {\bibinfo {author} {\bibfnamefont {D.}~\bibnamefont
  {Ahn}}, \bibinfo {author} {\bibfnamefont {T.}~\bibnamefont {Ralph}}, \ and\
  \bibinfo {author} {\bibfnamefont {R.}~\bibnamefont {Mann}},\ }\href
  {http://arxiv.org/abs/1008.0221} {\bibfield  {journal} {\bibinfo  {journal}
  {e-print arXiv:1008.0221v2 [quant-ph]}} (\bibinfo {year} {2010})}\BibitemShut
  {NoStop}%
\bibitem [{\citenamefont {Bennett}\ \emph {et~al.}(2009)\citenamefont
  {Bennett}, \citenamefont {Leung}, \citenamefont {Smith},\ and\ \citenamefont
  {Smolin}}]{BEN09}%
  \BibitemOpen
  \bibfield  {author} {\bibinfo {author} {\bibfnamefont {C.~H.}\ \bibnamefont
  {Bennett}}, \bibinfo {author} {\bibfnamefont {D.}~\bibnamefont {Leung}},
  \bibinfo {author} {\bibfnamefont {G.}~\bibnamefont {Smith}}, \ and\ \bibinfo
  {author} {\bibfnamefont {J.~A.}\ \bibnamefont {Smolin}},\ }\Doi
  {10.1103/PhysRevLett.103.170502} {\bibfield  {journal} {\bibinfo  {journal}
  {Phys. Rev. Lett.},\ }\textbf {\bibinfo {volume} {103}},\ \bibinfo {pages}
  {170502} (\bibinfo {year} {2009})}\BibitemShut {NoStop}%
\bibitem [{\citenamefont {Cavalcanti}\ and\ \citenamefont
  {Menicucci}(2010)}]{CAV10}%
  \BibitemOpen
  \bibfield  {author} {\bibinfo {author} {\bibfnamefont {E.~G.}\ \bibnamefont
  {Cavalcanti}}\ and\ \bibinfo {author} {\bibfnamefont {N.~C.}\ \bibnamefont
  {Menicucci}},\ }\href {http://arxiv.org/abs/1004.1219} { (\bibinfo {year}
  {2010})},\ \Eprint {http://arxiv.org/abs/arXiv:1004.1219 [quant-ph]}
  {arXiv:1004.1219 [quant-ph]} \BibitemShut {NoStop}%
\bibitem [{\citenamefont {Cavalcanti}\ \emph {et~al.}(2012)\citenamefont
  {Cavalcanti}, \citenamefont {Menicucci},\ and\ \citenamefont
  {Pienaar}}]{CAV12}%
  \BibitemOpen
  \bibfield  {author} {\bibinfo {author} {\bibfnamefont {E.}~\bibnamefont
  {Cavalcanti}}, \bibinfo {author} {\bibfnamefont {N.}~\bibnamefont
  {Menicucci}}, \ and\ \bibinfo {author} {\bibfnamefont {J.}~\bibnamefont
  {Pienaar}},\ }\href {http://arxiv.org/abs/1206.2725} {\bibfield  {journal}
  {\bibinfo  {journal} {e-print arXiv:1206.2725v1 [quant-ph]}} (\bibinfo {year}
  {2012})}\BibitemShut {NoStop}%
\bibitem [{\citenamefont {Ralph}\ and\ \citenamefont {Myers}(2010)}]{RAL10}%
  \BibitemOpen
  \bibfield  {author} {\bibinfo {author} {\bibfnamefont {T.~C.}\ \bibnamefont
  {Ralph}}\ and\ \bibinfo {author} {\bibfnamefont {C.~R.}\ \bibnamefont
  {Myers}},\ }\Doi {10.1103/PhysRevA.82.062330} {\bibfield  {journal} {\bibinfo
   {journal} {Phys. Rev. A},\ }\textbf {\bibinfo {volume} {82}},\ \bibinfo
  {pages} {062330} (\bibinfo {year} {2010})}\BibitemShut {NoStop}%
\bibitem [{\citenamefont {Ralph}\ \emph {et~al.}(2009)\citenamefont {Ralph},
  \citenamefont {Milburn},\ and\ \citenamefont {Downes}}]{RAL09}%
  \BibitemOpen
  \bibfield  {author} {\bibinfo {author} {\bibfnamefont {T.~C.}\ \bibnamefont
  {Ralph}}, \bibinfo {author} {\bibfnamefont {G.~J.}\ \bibnamefont {Milburn}},
  \ and\ \bibinfo {author} {\bibfnamefont {T.}~\bibnamefont {Downes}},\ }\Doi
  {10.1103/PhysRevA.79.022121} {\bibfield  {journal} {\bibinfo  {journal}
  {Phys. Rev. A},\ }\textbf {\bibinfo {volume} {79}},\ \bibinfo {pages}
  {022121} (\bibinfo {year} {2009})}\BibitemShut {NoStop}%
\bibitem [{\citenamefont {Penrose}(1998)}]{PEN98}%
  \BibitemOpen
  \bibfield  {author} {\bibinfo {author} {\bibfnamefont {R.}~\bibnamefont
  {Penrose}},\ }\Doi {10.1098/rsta.1998.0256} {\bibfield  {journal} {\bibinfo
  {journal} {Phil.Trans. R. Soc. Lon. Series A},\ }\textbf {\bibinfo {volume}
  {356}},\ \bibinfo {pages} {1927} (\bibinfo {year} {1998})}\BibitemShut
  {NoStop}%
\bibitem [{\citenamefont {Pikovski}\ \emph {et~al.}(2012)\citenamefont
  {Pikovski}, \citenamefont {Vanner}, \citenamefont {Aspelmeyer}, \citenamefont
  {Kim},\ and\ \citenamefont {Brukner}}]{PIK10}%
  \BibitemOpen
  \bibfield  {author} {\bibinfo {author} {\bibfnamefont {I.}~\bibnamefont
  {Pikovski}}, \bibinfo {author} {\bibfnamefont {M.~R.}\ \bibnamefont
  {Vanner}}, \bibinfo {author} {\bibfnamefont {M.}~\bibnamefont {Aspelmeyer}},
  \bibinfo {author} {\bibfnamefont {M.~S.}\ \bibnamefont {Kim}}, \ and\
  \bibinfo {author} {\bibfnamefont {C.}~\bibnamefont {Brukner}},\ }\Doi
  {10.1038/nphys2262} {\bibfield  {journal} {\bibinfo  {journal} {Nat. Phys.},\
  }\textbf {\bibinfo {volume} {8}},\ \bibinfo {pages} {393} (\bibinfo {year}
  {2012})}\BibitemShut {NoStop}%
\bibitem [{\citenamefont {Ralph}\ and\ \citenamefont {Downes}(2012)}]{RAL12}%
  \BibitemOpen
  \bibfield  {author} {\bibinfo {author} {\bibfnamefont {T.~C.}\ \bibnamefont
  {Ralph}}\ and\ \bibinfo {author} {\bibfnamefont {T.~G.}\ \bibnamefont
  {Downes}},\ }\Doi {10.1080/00107514.2011.640146} {\bibfield  {journal}
  {\bibinfo  {journal} {Contemporary Physics},\ }\textbf {\bibinfo {volume}
  {53}},\ \bibinfo {pages} {1} (\bibinfo {year} {2012})}\BibitemShut {NoStop}%
\bibitem [{\citenamefont {Wootters}\ and\ \citenamefont {Zurek}(1982)}]{WOO82}%
  \BibitemOpen
  \bibfield  {author} {\bibinfo {author} {\bibfnamefont {W.~K.}\ \bibnamefont
  {Wootters}}\ and\ \bibinfo {author} {\bibfnamefont {W.~H.}\ \bibnamefont
  {Zurek}},\ }\Doi {10.1038/299802a0} {\bibfield  {journal} {\bibinfo
  {journal} {Nature},\ }\textbf {\bibinfo {volume} {299}},\ \bibinfo {pages}
  {802} (\bibinfo {year} {1982})}\BibitemShut {NoStop}%
\bibitem [{\citenamefont {Walls}\ and\ \citenamefont {Milburn}(1994)}]{WAL94}%
  \BibitemOpen
  \bibfield  {author} {\bibinfo {author} {\bibfnamefont {D.~F.}\ \bibnamefont
  {Walls}}\ and\ \bibinfo {author} {\bibfnamefont {G.~J.}\ \bibnamefont
  {Milburn}},\ }\href@noop {} {\emph {\bibinfo {title} {Quantum Optics}}}\
  (\bibinfo  {publisher} {Springer-Verlag, Berlin},\ \bibinfo {year}
  {1994})\BibitemShut {NoStop}%
\bibitem [{\citenamefont {Kent}(2005)}]{KEN05}%
  \BibitemOpen
  \bibfield  {author} {\bibinfo {author} {\bibfnamefont {A.}~\bibnamefont
  {Kent}},\ }\Doi {10.1103/PhysRevA.72.012108} {\bibfield  {journal} {\bibinfo
  {journal} {Phys. Rev. A},\ }\textbf {\bibinfo {volume} {72}},\ \bibinfo
  {pages} {012108} (\bibinfo {year} {2005})}\BibitemShut {NoStop}%
\bibitem [{\citenamefont {Hanggi}\ and\ \citenamefont {Wehner}(2012)}]{HAN12}%
  \BibitemOpen
  \bibfield  {author} {\bibinfo {author} {\bibfnamefont {E.}~\bibnamefont
  {Hanggi}}\ and\ \bibinfo {author} {\bibfnamefont {S.}~\bibnamefont
  {Wehner}},\ }\href {http://arxiv.org/abs/1205.6894v1} {\bibfield  {journal}
  {\bibinfo  {journal} {e-print arXiv:1205.6894v1 [quant-ph]}} (\bibinfo {year}
  {2012})}\BibitemShut {NoStop}%
\bibitem [{Note1()}]{Note1}%
  \BibitemOpen
  \bibinfo {note} {One example of an appropriate metric can be found in \cite
  {POL94}.}\BibitemShut {Stop}%
\bibitem [{Note2()}]{Note2}%
  \BibitemOpen
  \bibinfo {note} {More precisely, we consider the time co-ordinate $\tau $ in
  the reference frame of the detectors, which are assumed to be at rest
  relative to each other. The spatial wave-packet on a given time-slice $\tau
  =t$ is denoted $G_i(\protect \textbf {x},t)$. If $(\protect \textbf
  {x}_{i},\tau ')$ is the detection event for the $i_{th}$ trajectory, then the
  temporal detector resolution is $G_i(\protect \textbf {x}_i,\tau -\tau
  ')\equiv G_i(\tau )$. See \cite {RAL09,PIE11} for details.}\BibitemShut
  {Stop}%
\bibitem [{\citenamefont {Pienaar}\ \emph {et~al.}(2011)\citenamefont
  {Pienaar}, \citenamefont {Myers},\ and\ \citenamefont {Ralph}}]{PIE11}%
  \BibitemOpen
  \bibfield  {author} {\bibinfo {author} {\bibfnamefont {J.~L.}\ \bibnamefont
  {Pienaar}}, \bibinfo {author} {\bibfnamefont {C.~R.}\ \bibnamefont {Myers}},
  \ and\ \bibinfo {author} {\bibfnamefont {T.~C.}\ \bibnamefont {Ralph}},\
  }\Doi {10.1103/PhysRevA.84.062316} {\bibfield  {journal} {\bibinfo  {journal}
  {Phys. Rev. A},\ }\textbf {\bibinfo {volume} {84}},\ \bibinfo {pages}
  {062316} (\bibinfo {year} {2011})}\BibitemShut {NoStop}%
\end{thebibliography}

\end{document}